# Intrusion Detection System Using Advanced Honeypots

Ram Kumar Singh
Deptt. of Information Technology
International School of Business & Tech.
Kampala, Uganda, East Africa.
E-mail: ramkumar.singh76@gmail.com
Mob: +256-715-141114

Prof. T. Ramanujam
Director.
Krishna Engg. College,
Mohan Nagar,Ghaziabad,U.P, India
E-mail: rj1944@gmail.com
Mob: +91 9999110885

*Abstract —* **As the number and size of the Network and Internet traffic increase and the need for the intrusion detection grows in step to reduce the overhead required for the intrusion detection and diagnosis, it has made public servers increasingly vulnerable to unauthorized accesses and incursion of intrusions[2]. In addition to maintaining low latency and poor performance for the client, filtering unauthorized accesses has become one of the major concerns of a server administrator.**

**Hence implementation of an Intrusion Detection System (IDS) [2] distinguishes between the traffic coming from clients and the traffic originated from the attackers or intruders, in an attempt to simultaneously mitigate the problems of throughput, latency and security of the network. We then present the results of a series of load and response time in the terms of performance and scalability tests, and suggest a number of potential uses for such a system.**

**The exponential growth of computer/network attacks are becoming more and more difficult to identify and the need for better and more efficient intrusion detection systems increases in step. The main problem with current intrusion detection systems is high rate of false alarms. Use of honeypots [1] provides effective solution to increase the security and reliability of the network.**

## I. INTRODUCTION

We present the design and implementation of a load balancer that distinguishes between the traffic coming from clients and the traffic originated from the attackers. This system is an attempt to simultaneously solve the problems of load balancing [3] and unauthorized intrusion.

If, in the process of forwarding requests, the balancer detects traffic as an attack on the server ('an exploit'), it is then directed to an alternative server - a type of honeypot. Conventional detection and forensics methodology can then be used to gather information on the intruder who will be unaware that they are not using a "real" server. Thus, the system will not only protect mission critical servers from unauthorized access in a manner transparent to the user, but allow for detailed data to be collected, which can later be used to take appropriate legal action against the incursion of intruders.

## II. BACKGROUND ON IDS

An intrusion detection system (IDS) inspects all inbound and outbound network activity and identifies suspicious patterns that may indicate a network or system attack from someone attempting to break into or compromise a system.

There are several ways to categorize an IDS:

**A. Misuse detection** vs. **anomaly detection**: in misuse detection, the IDS analyzes the information it gathers and compares it to large databases of attack signatures. Essentially, the IDS looks for a specific attack that has already been documented. Like a virus detection system, misuse detection software is only as good as the database of attack signatures that it uses to compare packets against. In anomaly detection, the system administrator defines the baseline, or normal, state of the network's traffic load, breakdown, protocol, and typical packet size. The anomaly detector monitors network[12] segments to compare their state to the normal baseline and look for anomalies.

**B.     Network-based** vs. **host-based systems**: in a network-based system, or NIDS, the individual packets flowing through a network are analyzed. The NIDS can detect malicious packets that are designed to be overlooked by a firewall's simplistic filtering rules. In a host-based system, the IDS examines at the activity on each individual computer or host.

**C.     Passive system** vs. **reactive system**: in a passive system, the IDS detects a potential security breach, logs the information and signals an alert. In a reactive system, the IDS responds to the suspicious activity by logging off a user or by reprogramming the firewall to block network traffic from the suspected malicious source.

Though they both relate to network security, an IDS differs from a firewall in that a firewall looks out for intrusions in order to stop them from happening. The firewall limits the access between networks in order to prevent intrusion and does not signal an attack from inside the network. An IDS evaluates a suspected intrusion once it has taken place and signals an alarm. An IDS also watches for attacks that originate from within a system.

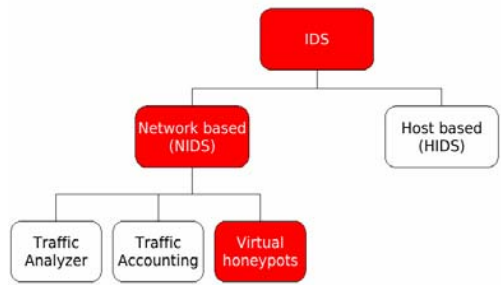

Figure 1: Intrusion Detection System hierarchy

## IV.   NETWORK      INTRUSION DETECTION SYSTEMS (NIDS)

Recent years have seen a drastic increase in   the popularity  of  Network   Intrusion Detection systems (NIDS). Intrusion Detection Systems (IDS) can be classified into two categories: Network-based Intrusion Detection Systems (NIDS) and Host-based Intrusion Detection Systems (HIDS). Existing          products have substantial network monitoring   capacity,  and   have been      largely successful at providing accurate reports on unauthorized activity. While these products have gotten fairly good at monitoring  and  reporting  on unusual activity, it has been stated, that the biggest problem with IDS  is "intelligently reacting to their output".

If an IDS detects an attack in progress, what action should be taken?. Furthermore, even if the organization deploying the IDS has the foresight create an attack response plan, if an administrator receives a page at 4:00 am, what are the chances that s/he will be able to react quickly enough to prevent the intrusion?  Under most existing systems chances for timely prevention of the intrusion highly depend  on  the  quick reaction of the human administrator.
All too often IDS logs are only consulted long after the damage from an attack has been done.

## III.      BACKGROUND ON HONEYPOTS

A honeypot is "an information system resource whose value lies in unauthorized or illicit use of that resource"[1].

A Honeypot is defined as "being a security resource whose value lies in being probed or compromised" [1]. They can be either host and/or network based, but are more often than not network based as all interaction is typically performed over a network connection. A honeypot's main utility comes from the fact that it simplifies the Intrusion Detection problem of separating "anomalous" from "normal" by having no legitimate purpose, thus any activity on a Honeypot can be immediately defined as anomalous. The characteristics of Honeypots make them well suited to the monitoring of malcious activity on networks, as well as being a valuable research tool in Computer Security.
Honeypot concepts are not particularly new; they can be traced back to early Computer Security papers such as Cliaord Stoll's Stalking the Wily Hacker [2]. However the study of Honeypots has recently been formalized. With this formalization has come a focus on research of attackers' methods and motivations [3] which has led to Honeypots being instrumental in the discovery of new security vulnerabilities [4].

## IV.    SOURCE DIRECT

Secure Direct is an attempt to address this problem: by providing a fully automated response to specific network intrusions,  it can eliminate the need for human decision making, and thus mitigate slow human response  times.

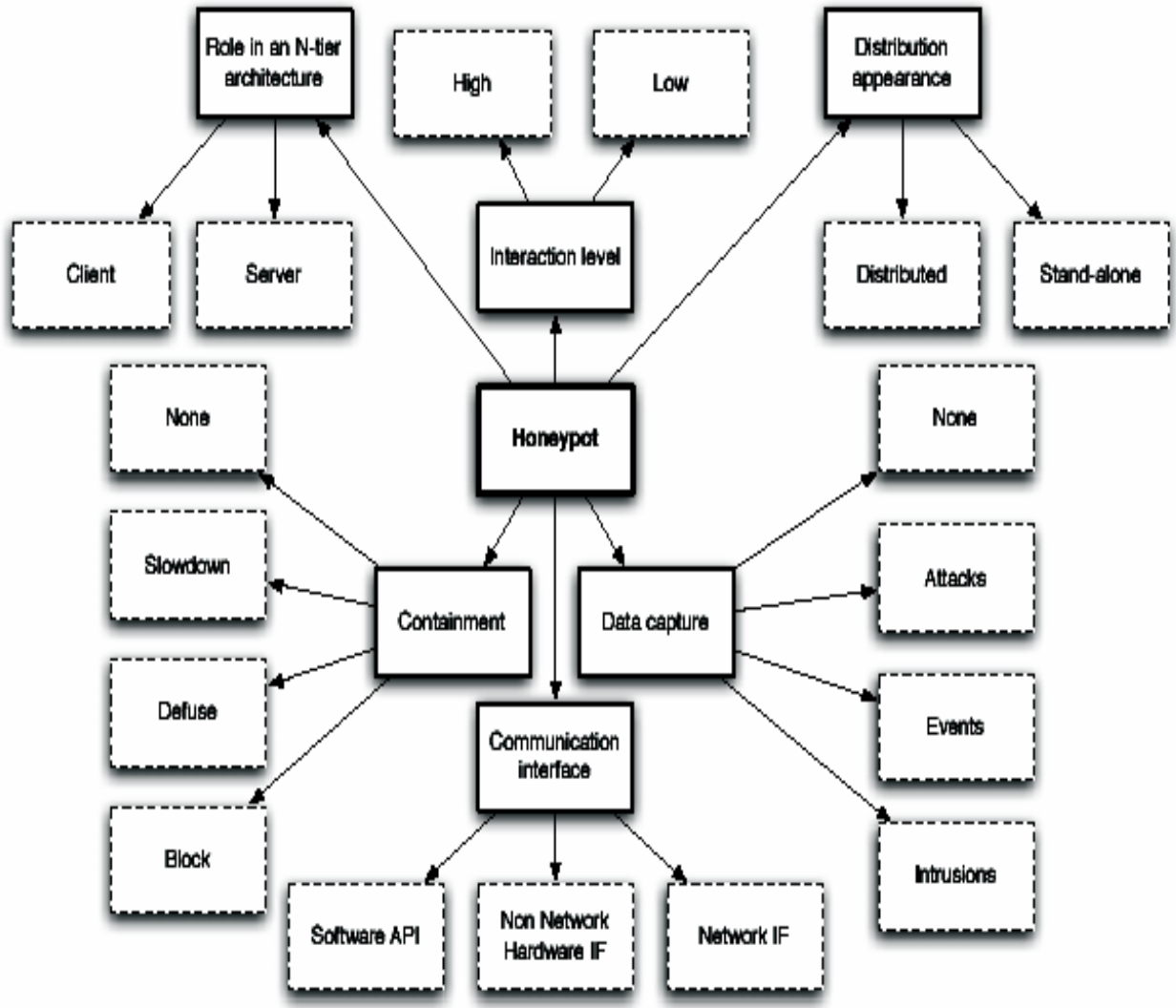

Figure2.  Taxonomy of Honeypot

## V.     ARCHITECTURE OF IDS USING ADVANCED HONYPOT

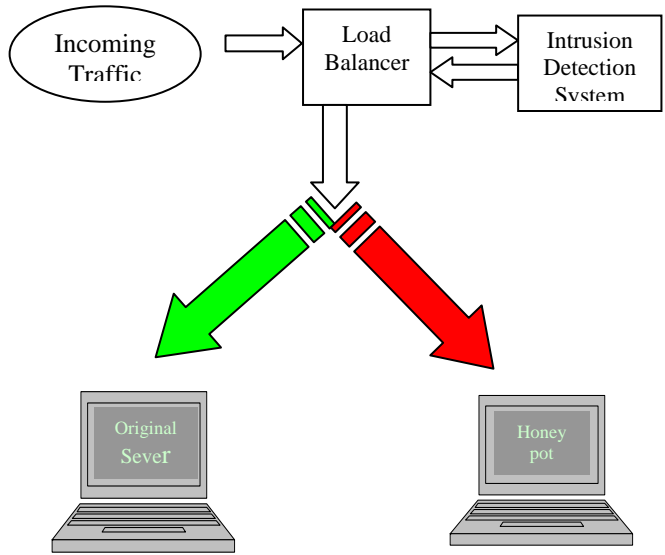

Figure 3. Architecture of IDS using Advanced Honypot

The activity at each process in the above diagram can be described   as follows:

1.  The load balancer [3] receives the request to the virtual IP address.    If the packet containing the    request has been fragmented, it is reassembled.

2.  The Load balancer opens a TCP connection to the IDS Process, and sends the content of the packet (less the headers) over that connection.

3.  The IDS process checks the content of the   packet against its database of known attacks, and returns a Boolean result to the load balancer over the same TCP connection.

4.  On receiving the result, the load balancer c l o s e s the TCP connection. If the result from the IDS was "true" (Indicating the presence of an attack) the packet is forwarded   to   the Honeypot. Otherwise, a server is selected from the active server pool in a round-robin fashion and the packet is forwarded to the server.

The    design    of    this system    entailed overcoming a number of    challenges. What these challenges were, and how they were addressed    is    discussed in the remainder of this section.

Honeypots are a highly flexible security tool with different applications for security. They don't fix a single problem. Instead they have multiple uses, such as prevention, detection, or information gathering. Honeypots all share the same concept, a security resource that should not have any production or authorized activity. In other words, deployment of honeypots in a network should not affect critical network services and applications. A honeypot is a security resource whose value lies in being probed, attacked, or compromised. There are two general types of honeypots: Production honeypots are easy to use, capture only limited information, and are used primarily by companies or corporations; and Research honeypots are complex to deploy and maintain, capture extensive information, and are used primarily by research, military, or government organizations.

http://www.honeypots.net Honeypots are increasingly used to provide early warning of potential intruders, identify flaws in security strategies, and improve an organization's overall security awareness. "Honeypots can simulate a variety of internal and external devices, including Web servers, mail servers, database servers, application servers, and even firewalls. As a software development manager, I regularly use honeypots to gain insight into vulnerabilities in both the software my team writes and the OS upon which we depend."

http://www.windowsitpro.com/Windows/Article/ArticleID/44711/44711.html A honeypot is a security resource whose value lies in being probed, attacked, or compromised. This means that whatever we designate as a honeypot, it is our expectation and goal to have the system probed, attacked, and potentially exploited. A honeypot also is a detection and response tool, rather than prevention which it has a little value in. (http://www.securitydocs.com/library/2692) A better way to think of a honeypot is as an Internet-attached server that acts as a decoy, luring in potential hackers in order to study their activities and monitor how they are able to break into a system.

Honeypots are designed to mimic systems that an intruder would like to break into but limit the intruder from having access to an entire network. If a honeypot is successful, the intruder will have no idea that s/he is being tricked and monitored. Most honeypots are installed inside firewalls so that they can better be controlled, though it is possible to install them outside of firewalls. A firewall in a honeypot works in the opposite way that a normal firewall works: instead of restricting what comes into a system from the Internet, the honeypot firewall allows all traffic to come in from the Internet and restricts what the system sends back out. By luring a hacker into a system, a honeypot serves several purposes: The administrator can watch the hacker exploit the vulnerabilities of the system, thereby learning where the system has weaknesses that need to be redesigned. The hacker can be caught and stopped while trying to obtain root access to the system.By studying the activities of hackers, designers can better create more secure systems that are potentially invulnerable    to    future    hackers. (http://www.securitydocs.com/library/2692)

## VI.  LOAD BALANCING

In designing   a load balancer [3] for Secure Direct we   have three main focuses, namely,   to provide

High-Availability by handling hardware failure in the web-cluster, maintain high speed access to the cluster, and ensure the balancer itself does not become a security hole. High availability is achieved by simply pinging the servers at regular intervals, and removing them from the server pool if no response is received.

The second challenge is to secure the load balancer. Our strategy is to protect it from any irrelevant traffic. Only traffic to a specific port on the virtual IP will be processed by the load balancer and any other malicious access is simply ignored. The load balancer uses a technique known as 'Proxy-ARP' to respond to ARP requests from the router to the virtual IP address.

The web servers have their loop back interface3 configured with the virtual IP address, but are set not to respond to ARP packets. This, at the network layer, there is no way to tell which servers are configured with the virtual IP.

The load balancer process is implemented as a multi-threaded process in Java. The main thread is responsible for reading packets off the wire and if they are destined to virtual IP and their destination port is our desired service port, it hands them to the control thread. The control thread then communicates with the IDS process and decides to pass the packet to the production servers or direct it to the Honeypot.

In case of TCP applications, after redirecting a request to a server, the load balancer process should hold the information about which request is forwarded to which server in order to forward all of the ACK packets in a particular session. We keep the information about each connection in a table. As the load balancer only passes the traffic from client to server, it is unable to see the last FIN sent by the server, and therefore there is no way that the load balancer can find out when the conversation between two hosts is over. Therefore we define a time stamp for each connection. Each time a packet is received on a connection its time stamp is updated. The connections will be removed from the table if its time stamp is not updated for a certain amount of time (which has been set at arbitrarily at 4 minutes during our tests).

## VII. INTRUSION DETECTION

Internally, the implementation of the IDS portion of Secure Direct is very simple. The IDS process runs as a concurrent TCP server, and listens for client requests on a specified port.

When a connection is made, the IDS fork a child process when receives the content of the packet, and then checks it against a database of known attacks.

It then returns the result of this check to the load balancer process, which makes the decision on whether to forward the request to the production server cluster, or the Honeypot.

The multithreaded design of the load balancer ensures that multiple requests from a client will not get 'mixed up'; however, it is possible that an attack would occur from a single IP at the same time as a valid request. In this case, the initial, harmless packets may be safely forwarded to the real servers until the IDS process finds the attack-packet and detects the signs of intrusion. At this point, it immediately informs the load balancer process to discontinue forwarding packets to the real server, and to send an RST packet to the corresponding server to end the connection. Thus, the server will never receive the attack. In the attacker side, observing silence from the server side causes it to assume the server has crashed and possibly causes it to try to re-connect. However from this point, after detecting the intrusion, all the incoming traffic from the attacker's IP will be forwarded to the Honeypot.

One of the key points in maintaining speed of this system is that, once an IP is recognized as an Attacker IP, packets coming from that source are not passed to the IDS process any more. The load balancer process directs the incoming traffic from attackers to the Honeypot.

## VIII. THE TCP LAYER

Secure Direct is implemented for applications that use TCP/IP as their transport protocol. Secure Direct is designed (like any good security product) to be failed-closed system, which means if it crashes, it is no longer possible to access either web server through the virtual IP. Consequently the attackers can not crash Secure Direct and access the unprotected system.

One of the major challenges for Secure Direct is to deal with attackers who try to fool the system by forcing it to analyze the packets inconsistent with what is received in the end-system. In the cases that the intrusion detection system runs at the different host-OS than the end-system, this can be a serious concern. The attacker can take advantage of differences between the IDS and the end-system in dealing with the packets that do not fully comply with the standard protocols, and send some packets that are discarded by the end-host but accepted by the IDS, or vice versa. If Secure Direct uses TCP/IP, the inconsistency between the IDS and the end-host may appear in IP level or TCP level [5]. TCP protocol uses sequence numbers to preserve the order of the incoming packets. The end-system waits until it receives all the sequence numbers required for re-assembling the data. If there is a missing sequence number, the end-system will not accept the consequent packets and waits until it receives the packet with the sequence number it is waiting for. Therefore one way to try to fool the system is to send two packets with the same sequence numbers, one containing false data to be accepted by the IDS and discarded by the end-host, and the other one containing the attackers desired data to be accepted by the end host, and skipped by the IDS. Thus, whenever it detects two different packets with the same sequence number (and a sane checksum) for one connection, it considers it as an attack and prevents the load balancer from sending that packet to the end-host. Furthermore it marks the source IP of this packet, as an attacker IP, causing the load balancer to forward all the consequent packets originated from this IP to the Honeypot.

An alternative way to break into the system is using IP fragmentation, hoping that IDS and the end-host follow two different methods for re-assembling IP fragments. Secure Direct however re-assembles the IP fragmented packets in the load balancer and forwards the assembled packet to the end-host. Therefore what IDS analyzes is completely consistent with what end-host sees. This type of implementation should drastically reduce the chances of an attacker breaking into the system.

## IX.    EXPERIMENTAL RESULTS

I chose Windows 2003 Server and professional systems with a 2.0GHz processor and 512 MB of RAM with a CDROM drive. Windows 20003 was the best choice to since it can be secured the most from the operating systems I had available to chose from, Windows XP, Windows 2000 Server and Windows 2003 Professional. .To archives the findings for evidence if needed and also to see if a pattern would develop over time to the probability of an attack and what type of attack.

The experimental results we describe the detail of this experiment conducted with network simulator NS2 and configuration of a system 3.0GHz processor and 512 MB of RAM.

### TEST ENVIRONMENT

A simple test environment requires a minimum of 3 servers and 4 IP addresses. Each server is assigned a single IP address for basic TCP/IP connectivity, while the fourth is used as a virtual IP, which will be accessed by clients, and balanced by the balancer to either of the two servers depending on the contents of the client request.

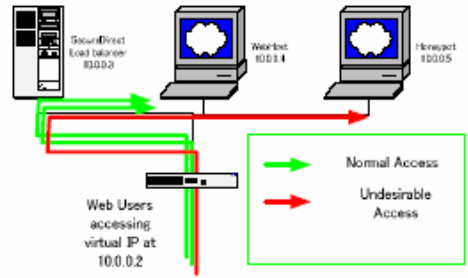

Figure 4. Test Network Environment

Some intrusion correlation systems do not use a raw data stream (like network or audit data) as input, but instead rely upon alerts and aggregated information reports from IDSs and other sensors [20, 21]. We need to develop systems that can generate realistic alert log files for testing correlation systems. A solution is to deploy real sensors and to "sanitize" the resulting alert stream by replacing IP addresses. Sanitization in general is difficult for network activity traces but it is relatively easy in this special case since alert streams use well defined formats and generally contain little sensitive data (the exception being IP addresses and possibly passwords). Alternately, some statistical techniques for generating synthetic alert datasets from scratch are presented by [16].

IDS testing efforts vary significantly in their depth, scope, methodology, and focus. The characteristics of some of the more developed efforts listed in chronological order. This demonstrates that evaluations have increased in complexity over time to include more IDSs and more attack types, such as stealthy and denial of service (DoS) attacks. Only research evaluations have included novel attacks designed specifically for the evaluation, evaluated the performance of anomaly detection systems, and generated curves. Evaluations of commercial systems have included measurements of performance under high traffic loads. Traffic loads were generated using real high-volume background traffic mirrored from a live network and also with commercial load testing tools. The remainder of this section provides details concerning all the evaluations, as well as for a few other more limited evaluations. Evaluations of research systems are described first, followed by descriptions of commercial system evaluations.

The load balancer itself doesn't gather any information internally, the addition of a Network Based Intrusion Detection System (NIDS) on the network can be used to perform this task, while the intruder is harmlessly attacking the Honeypot server.

However, some difficulties exist when using this approach:

1. Sanitization attempts may end up removing much of the content of the background activity thus creating a very unrealistic environment.

2. Sanitization attempts may fail causing an accidental release of sensitive data. This scenario is very possible since it is infeasible for a human to verify the sanitization of a large volume of data. We feel that this risk is one that most organizations will not tolerate for the sake of a research project.

3. Since attacks have to be injected somewhat artificially into the sanitized data stream (regardless of the method used), the attacks will not realistically interact with the background activity. For example, buffer overflow attacks may be launched against a web server and cause the server to crash, but normal background requests to the web server may continue. This lack of interaction between the attacks and the background activity could be a problem when testing IDSs.

4. When sanitizing real traffic, it may be difficult to remove attacks that existed in the data stream. If one is merely testing hit rates, then having unidentified attacks in the data is an inconvenience but not a major problem. However, having unidentified attacks in the background activity would pose a problem for any false positive testing. Further, sanitizing data may remove information needed to detect attacks.

## RESULTS

While the basic functionality of a content-based load balancer is relatively easy to achieve, such a system is only useful to the extent that it is scaleable and stable under load. At a modest load, the balancer showed very stable performance:

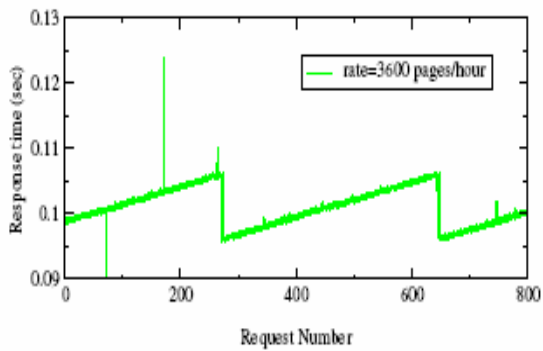

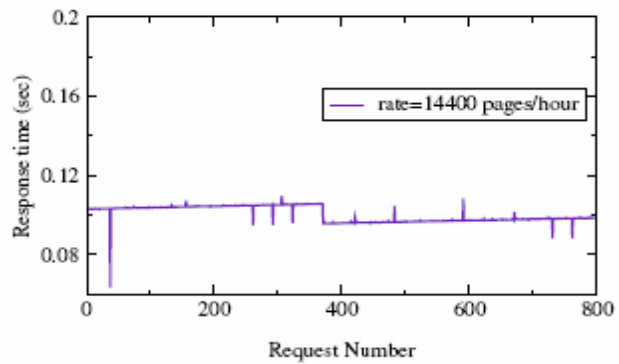

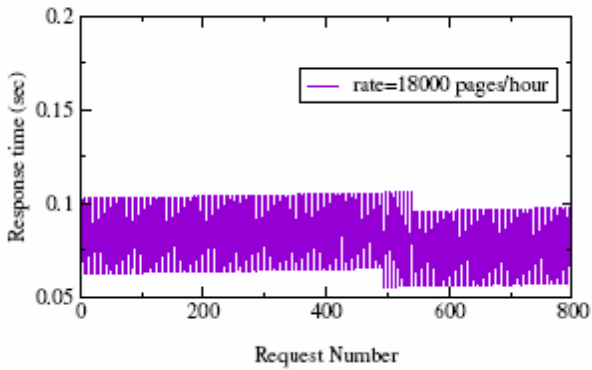

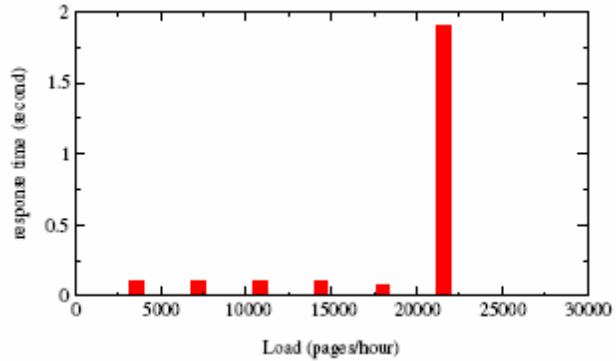

# CONCLUSIONS

In this work, we have described some of the previous efforts to measure IDS, and we have outlined some of the difficulties that have been encountered. We believe that a periodic, comprehensive evaluation of IDSs could be valuable for network managers, information security officers and data managers. However, because both normal and attack traffic are so variable from site to site, and because normal and attack traffic evolve over time

Solving the problem of high availability and security simultaneously offers the opportunity for more reliability than systems which solve the problems separately, in addition to being easier to implement, and offering increased opportunity for recording and data analysis.

The integration of these two technologies, however, is a non-trivial task. A fine balance must be achieved between speed and robustness of IDS features – it is equally bad to have the web server crash because an attack was missed, or drop large amounts of traffic due to an overly through IDS becoming a bottleneck. Additionally, while a content-based load balancer offers many advantages over separate Load Balancing and IDS solutions, it also suffers from the drawbacks of both.

It has been said that the three things most important to a server manger are Security (the ability to withstand hacking), High-Availability (the ability to withstand hardware failure), and low-latency (the ability to service requests quickly). With the current implementation of secure direct, the first two criteria, security and high availability, are satisfied. In regards to latency, our experiments have (once again) shown that there is a trade-off between security and performance. While improving the efficiency of the programming can mitigate this problem to an extent, the trade-off is certain to remain, and choosing the right balance will likely continue to be a difficult task for the system administrator. It is the authors' opinion that this type of system could provide a much needed additional security tool, however, a good deal of streamlining work remains before it could be widely deployed.

## ACKNOWLEDGMENT

The Success of this research work would have been uncertain without the help and guidance of a dedicated group of people. I would like to express my true and sincere acknowledgements as the appreciation for their contributions. I also express my sincere thanks to Dr. Ajay Sharma, Director General, KIET, Ghaziabad, India and Prof. Verges Mudamattam, Director, ISBAT, Kampala, Uganda, East Africa, for their encouragement and support.

## REFERENCES

[1]   The Honeynet Project. Know Your Enemy : Honeynets (May 2005) http://www.honeynet .org/papers/honeynet/.

[2]   Cliaord Stoll. Stalking the Wily Hacker. Communicationsof the ACM. pp 484- 497. 1988.

[3]   Honeynet Research Alliance. Project Honeynet Website. Retrieved May 16th 2003 from the World Wide Web: http://project.honey.org

[4]   Computer Emergency Response Team. dtscpd Exploit Advisory. Advisory CA-2002-01 Exploitation of Vulnerability in CDE Subprocess

[5]   Balance, An open Source Load Balancing TCP Proxy: http://balance.sourceforge.net

[6]   Big/ip: http://www.bigip.com/bigip/

[7]   [E. Balas and C. Viecco, "Towards a Third Generation Data Capture Architecture of Honeypots," Proceedings of the 6th IEEE information Assurance Workshop, West Point (IEEE, 2005)

[8]   Bruce Schnier, Decrets and Lies: Digital Security in a Networked World, Wiley Publications, 2005

[9]   Hogwash: http://hogwsh.sourceforge.net/

[10]  Honeypot Project: http://www.honeypot.org

[11]  ManTrap: http://www.recourse.com

[12]  Martin Roesch, Snort- Ligthweight Intrusion Detection  for Networks, Proceedings  of LISA'99 : 13th System Administration Conference, Seattle, Washington USA, 2005

[13]  Pen A load balancer for simple tcp based protocols such as http or smtp for Unix: http://siag.nu/pen/

[14]  Thomas Ptacek, Thimoty N. Newsham, Insertion, Evasion, and Denial of Service: Eluding Network Intrusion Detection, Technical Report, Secure Networks Inc. 2000

[15]  The Honeynet Project. Know Your Enemy : Sebek (November 2003) http://www.honeynet .org/papers/sebek.pdf/.

[16]  Haines J.A., Rossey L.M., Lippmann R.P., and Cunningham R. K.. *Extending the DARPA Off-Line Intrusion Detection Evaluations* in *Darpa Information Survivability Conference and Exposition (DISCEX) II*. 2001. Anaheim, CA: IEEE Computer Society.

[17]  Lance Spitzner. Honeypots: Tracking Hackers. Addison-Wesley, Boston. 2002.

[18]  Kinsella, J. (January 2005). Build a PC Honeypot. Retrieved November 5, 2005, from Windows IT Pros Website: http://www.windowsitpro.com/Windows/Article/ArticleID/44711/44711.html

[19] Noordin, M. (November 5, 2004). Honeypots Revealed. Retrieved November 5, 2005, from Security Docs.com Website: http://www.securitydocs.com/library/2692

[20] Vigna G., Kemmerer R.A., Blix P. "Designing a Web of Highly-Configurable Intrusion Detection Sensors", Recent Advances in Intrusion Detection (RAID 2001), Davis, CA, October 2001.

[21] Valdes A., and Skinner K. "Probabilistic Alert Correlation", Recent Advances in Intrusion Detection (RAID 2001), Davis, CA, October 2001.

AUTHORS PROFILE

Authors Profile … **Ram Kumar Singh**[ M.Tech.(Comp. Sci. & Engg.), B.Engg. (Electronics and Communication Engg.), CCNA, Polyt Echnichnic Diploma in Computer Science] is Senior Lecturer and CISSP Trainer in Department of Information Technology, International School of Business and Technology, Kampala,Uganda. His research area is Network Security and Information Security.

**Prof. T. Ramanujam** [BE(Hons) in Electrical Engg, M Tech (Electrical Engg with specialisation in Computer Science from IIT, Kanpur) and M Sc in Military History from Madras University] is the (Founder) Director of Krishna Engineering College, Mohan Nagar, Ghaziabad, U.P., India.. He retired as an Air Vice Marshal of Indian Air Force after 35 years of distinguished service and is in the field of academics for the last 8 years. He was Dean (Academics), KIET, Ghaziabad for 2 years. He is currently doing Ph D in areas related to computational Biology and Bioinformatics.